\documentclass[conference]{IEEEtran}

\pdfoutput=1

\usepackage{caption}
\usepackage{subcaption}
\usepackage{lipsum}
\usepackage{graphicx}
\usepackage{cite}
\usepackage{comment}
\usepackage{pgf}
\usepackage{amssymb}
\usepackage{amsthm}
\usepackage{amsfonts}
\usepackage{amsmath}
\usepackage{epsfig}
\usepackage{color}
\usepackage{fancybox}
\usepackage{textcomp}
\usepackage{multirow}
\usepackage{setspace}
\usepackage{psfrag}
\usepackage[bookmarks=false]{hyperref}
\usepackage[ruled,vlined,linesnumbered]{algorithm2e}
\usepackage{booktabs}

\usepackage{tabularx}
\newcolumntype{L}[1]{>{\raggedright\arraybackslash}p{#1}}
\newcolumntype{C}[1]{>{\centering\arraybackslash}p{#1}}
\newcolumntype{R}[1]{>{\raggedleft\arraybackslash}p{#1}}

    \def\Complex{{\rm\rule[.23ex]{.03em}{1.1ex}\kern-.3em{C}}}

    \newcommand{\be}{\begin{equation}} \newcommand{\ee}{\end{equation}}
    \newcommand{\bea}{\begin{eqnarray}} \newcommand{\eea}{\end{eqnarray}}
    \newcommand{\benum}{\begin{enumerate}} \newcommand{\eenum}{\end{enumerate}}

    \newcommand{\qf}{{\bf f}}

    \newcommand{\qn}{{\bf n}}
    
    \newcommand{\qp}{{\bf p}}
    
    \newcommand{\qr}{{\bf r}}

    \newcommand{\qu}{{\bf u}}
    
    \newcommand{\qw}{{\bf w}}

    \newcommand{\qz}{{\bf z}}

    \newcommand{\qB}{{\bf B}}

    \newcommand{\qH}{{\bf H}}

    \newcommand{\qL}{{\bf L}}

    \newcommand{\qP}{{\bf P}}
    
    \newcommand{\qR}{{\bf R}}

    \newcommand{\qlambda}{{\boldsymbol \lambda}}

    \newcommand{\qtheta}{{\boldsymbol \theta}}

    \newcommand*{\argmax}{\operatornamewithlimits{argmax}\limits}

\begin{document}

\title{Local Cyber-physical Attack with Leveraging Detection in Smart Grid}

\author{\IEEEauthorblockN{
Hwei-Ming Chung\IEEEauthorrefmark{1}, Wen-Tai Li\IEEEauthorrefmark{2}, Chau Yuen\IEEEauthorrefmark{2}, Wei-Ho Chung\IEEEauthorrefmark{1}, and Chao-Kai Wen\IEEEauthorrefmark{3}}

\IEEEauthorblockA{\IEEEauthorrefmark{1} Research Center for Information Technology Innovation, Academia Sinica, Taipei, Taiwan 115}
\IEEEauthorblockA{\IEEEauthorrefmark{2} Engineering Product Development, Singapore University of Technology and Design, Singapore 487372}
\IEEEauthorblockA{\IEEEauthorrefmark{3} Institute of Communications Engineering, National Sun Yat-sen University, Kaohsiung, Taiwan 804}

\IEEEauthorblockA{Emails: \IEEEauthorrefmark{1}\{hweiming.chung, whc\}@citi.sinica.edu.tw,\\ \IEEEauthorrefmark{2}\{wentai\_li, yuenchau\}@sutd.edu.sg \IEEEauthorrefmark{3}chaokai.wen@mail.nsysu.edu.tw}
}

\maketitle
\begin{abstract}
A well-designed attack in the power system can cause an initial failure and then results in large-scale cascade failure. 
Several works have discussed power system attack through false data injection, line-maintaining attack, and line-removing attack. However, the existing methods need to continuously attack the system for a long time, and, unfortunately, the performance cannot be guaranteed if the system states vary.
To overcome this issue, we consider a new type of attack strategy called combinational attack which masks a line-outage at one position but misleads the control center on line outage at another position.  
Therefore, the topology information in the control center is interfered by our attack.
We also offer a procedure of selecting the vulnerable lines of its kind.
The proposed method can effectively and continuously deceive the control center in identifying the actual position of line-outage. 
The system under attack will be exposed to increasing risks as the attack continuously. 
Simulation results validate the efficiency of the proposed attack strategy.
\end{abstract}

\section{Introduction}
\IEEEPARstart{P}{ower} system plays an important role in supporting the modern economy.
Initial failures in power system, without being promptly detected, may lead to large-scale cascade failure, and have adverse affects on nation's economy and security \cite{cascade-failure}.
Therefore, in the operation control center, various data processing modules such as state estimation (SE) and bad data detection are built to prevent the system operation from failures and malicious attacks.
Although many protection and detection methods are used in system operation, these mechanisms can be corrupted by injecting carefully predesigned data to the measurements sent by Supervisory Control and Data Acquisition (SCADA).
The topic has attracted much attention in the past few years \cite{2009-liu-FDI,2012-M-FDI,2015-zli-FDI, 2015-sankar-mask-outage, 2014-zli-mask-outage,2016-zli-mask-outage-local, 2016-sankar-mask-outage, 2013-kim-topology-attack,2016-zli-topology-attack}.

In \cite{2009-liu-FDI}, the authors proposed the classic false data injection (FDI) attacks that can avoid being detected by existing bad data detection techniques if an attacker has the ability to alter the measurements of sensors and capture sufficient knowledge of the power system.
Such FDI attacks are also known as \emph{cyber attacks}.
The designed attacks should obey the physical laws (Kirchhoff's Current Law, KCL, and Kirchhoff's Voltage Law, KVL).
The authors in \cite{2012-M-FDI} and \cite{2015-zli-FDI} studied the classic FDI attacks with incomplete information of the system, and  \cite{2012-M-FDI} and \cite{2015-zli-FDI} revealed that the attacks have the ability of passing the SE and bad data detection with only reduced network information.

Another type of attack called \emph{cyber-physical attacks} involving cyber and physical levels have been investigated which can more efficiently interfere the operation of the system compared to classic FDI attack with only pure cyber attacks.
For example, there are two types of cyber-physical attacks, which are line-removing attack and line-maintaining attack as described in \cite{2015-sankar-mask-outage}.
The line-maintaining attacks mean that an attacker can let the target line be physically disconnected, and simultaneously mask this outage event with the altered measurements of sensors.
The other advanced line-maintaining attacks have been studied in \cite{2014-zli-mask-outage,2016-zli-mask-outage-local,2016-sankar-mask-outage}.
Specifically, the authors masked the outage event with local redistribution attack and extended to attack with incomplete topology information \cite{2014-zli-mask-outage,2016-zli-mask-outage-local}.
The attack model was further derived with power flow method \cite{2016-sankar-mask-outage}.

The line-removing attack is that an attacker generates a fake outage event so as to disturb the regular system operation.
The attack has to avoid the trivial solution; otherwise, it can be easily detected by the control center.
With this approach, the attacker can mislead the control center with an incorrect network topology and then make the system into unstable situation due to wrong dispatches.
The line-removing attacks have been studied with partial and whole information of the system, and mitigated with the countermeasure for the proposed attack \cite{2013-kim-topology-attack}.
The authors of \cite{2016-zli-topology-attack} focused on the line-removing attack in the local area, and proposed the method of finding the attack region.
While implementing this attack, one must notice that not all transmission lines in the power system can be selected as attack targets because some lines are strictly protected by the control center.
Only few studies, such as \cite{2015-sankar-mask-outage}, considered the rules for selecting target lines.

Based on the discussions above, the previous approaches have obtained promising results and demonstrated the potential of the \emph{cyber-physical attacks}.
However, there is no guarantee that the line-maintaining attacks are always unobservable.
To this end, the concept of the line-removing attack may be applied simultaneously to fake an obvious outage in order to attract the attention of control center, so that the disconnected line has lower chance to be identified.
Additionally, with this approach, the longer the control center in figuring the problem at fake outage positions, the more risky the system is.

Inspired by the above observations, we develop a novel attack strategy that combines the line-removing and line-maintaining attack strategies.
The attack is implemented in the local area and cannot be detected easily because our design makes sure that the physical laws of the power system are satisfied.
In addition, unlike previous studies which randomly select the target lines, a rule of deciding the target lines is proposed in this work.
To this end, we employ the line outage distribution factors (LODFs) as the impact of the attack line selections.
The contributions of this study are as follows:
\begin{itemize}
\item We propose a novel attack strategy called combinational attack whose goal is to attack the transmission line and simultaneously mask the real outage event with misleading the control center into another fake outage line.

\item We design a selection rule based on LODF for selecting the target lines instead of random selection.

\item To mislead the control center, the corresponding power flow must be dispatched according to the pattern of target line and misleading line.
Hence, we propose an algorithm based on breadth-first search (BFS) \cite{BFS_algo}, which is generally used for searching graph structure.

\item To test the effectiveness of the proposed attack strategy, the conventional SE and bad data detection are applied.
The simulation results reveal that the misleading line can be actually detected by control center and the real outage event can be successfully hidden at the same time.

\end{itemize}

\section{System Model}\label{sec:system_model_problem_formulation}
\begin{figure}
\begin{center}
\resizebox{3.5in}{!}{%
\includegraphics*{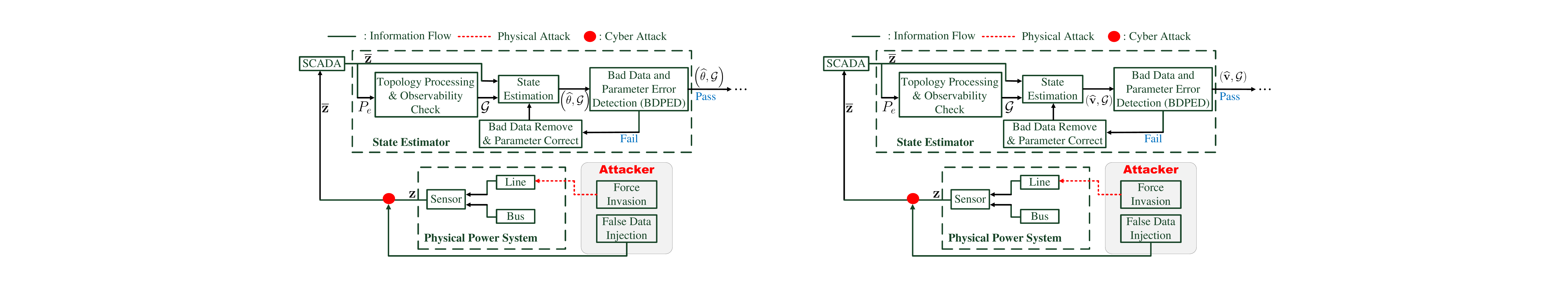} }%
\caption{The system block diagram}\label{fig:system_model}
\end{center}
\end{figure}

The system considered in this study is shown in Fig. \ref{fig:system_model}, which is divided into two parts, including state estimator, and cyber-physical attack model.
In this section, we briefly illustrate the state estimator based on the DC model, and then the proposed attack strategy will be introduced in next section.

\subsection{DC Power Flow Model}
We consider a power transmission network with $n_{b}$ buses and $n_{br}$ lines, and let $\mathcal{N}$ and $\mathcal{E}$ respectively be the sets of buses and lines.
The power network can then be represented as a graph denoted as $\mathcal{G}=\{ \mathcal{N}, \mathcal{E}\}$.
Assuming a line $l \in \mathcal{E}$ that connects bus $i$ and $j$, and then the power of line flowing from bus $i$ to $j$ denoted as $P_{l}$ can be represented as
\begin{equation}\label{eq:flow_eq}
P_{l} = \frac{\theta_{i} - \theta_{j}}{x_{l}},
\end{equation}
where $x_{l}$ is the reactance of line $l$, and $\theta_{i}$ and $\theta_{j}$ are the phases of bus $i$ and $j$, respectively.
With (\ref{eq:flow_eq}), the vector of all power flows $\qp = [P_{1} \cdots P_{n_{br}}] \in \mathbb{R}^{n_{br} \times 1}$ and the phase angles of the buses $\qtheta = [\theta_{1} \cdots \theta_{n_{b}}] \in \mathbb{R}^{n_{b}\times 1}$ should satisfy
\begin{equation}\label{eq:DC_model}
\qp = \qB_f \qtheta ,
\end{equation}
where $\qB_{f} \in \mathbb{R}^{n_{br}\times n_{b}}$ is a matrix whose row indicates the corresponding line, and the column presents the direction of line's flow.
Therefore, the $l$-th row of $\qB_{f}$ which represents line $l$ flowing from bus $i$ to bus $j$ can be formulated as
\begin{equation}
B_{f \{ l, k\}} = \left\{
\begin{array}{ll}
                \frac{1}{x_{l}},      & \mbox{if}~~ k = i,  \\
                \frac{-1}{x_{l}},     & \mbox{if}~~ k = j,  \\
                0,                    & \mbox{others}.
\end{array}
        \right.
\end{equation}

\subsection{Linear State Estimation}\label{sec:SE}
Based on the DC power flow model, the system states are phase angles, $\qtheta$, and therefore the measurements received by SCADA system without attack can be expressed as
\begin{equation}
\qz = \qH(P_{e}, \mathcal{G}) \qtheta+ \qn.
\end{equation}
Here, $\qz$ commonly comprises of the measurements of bus injection power and line power flow, and then $\qH$ is the jacobian matrix which depends on the network topology $\mathcal{G}$ and network parameter vector $P_{e}\in \mathbb{R}^{n_{br} \times 1}$ representing the parameter errors.
$\qn$ is the measurement errors.
We further denote the measurements modified by the attacker with $\overline{\qz}$.

With the measurement expression, we adopt weighted least-squares (WLS) SE to estimate the system state $\qtheta$.
The objective of the SE problem is to minimize the sum of the squares of the weighted deviations of the estimated measurements from $\overline{\qz}$.
The SE problem is then solved by the following optimization problem with assumption of zero parameter errors
\begin{subequations} \label{eq:SE_WLS}
\renewcommand\minalignsep{0em}
\begin{align}
\mathcal{F}1 :  & \min_{\widehat{\qtheta}}  \left( \overline{\qz} - \qH\left(P_{e}, \mathcal{G}\right)  \widehat{\qtheta}  \right)^{T} \qR^{-1} \left( \overline{\qz} - \qH\left(P_{e}, \mathcal{G}\right)  \widehat{\qtheta} \right)  \\
& \mbox{s.t.}   ~~~P_{e} = 0,
\end{align}
\end{subequations}
where $\widehat{\qtheta}$ is the estimated system state, $P_{e}$ is the parameter error vector, $\qR $ is the measurement error covariance matrix.

\subsection{Bad Data and Parameter Error Detection}\label{sec:bad_data_detec}
After applying SE, we have to pass through the bad data and parameter error detection to ensure there is no bad data or parameter errors within the measurements.
In this context, the normalized residual and parameter error method is employed for detection.

The measurement residual vector can be represented as
\begin{equation}
\qr =  \overline{\qz} - \qH\left(P_{e}, \mathcal{G}\right)  \widehat{\qtheta} .
\end{equation}
If the Lagrangian multiplier method is applied in (\ref{eq:SE_WLS}), $\qlambda$ is the Lagrangian multiplier related to the parameter error.
Given $\qr$ and $\qlambda$, the normalized residual $\qr^{N}$ and normalized parameter errors $\qlambda^{N}$ can be calculated.
The normalized residuals are linked to the corresponding measurements, and the normalized parameter errors are related to the corresponding line's parameter.
References \cite{2006-Abur-parameter-SE,2016-Abur-parameter-SE,abur-book} provide further details.
With the $\qr^{N}$ and $\qlambda^{N}$, the errors are regarded as Gaussian distribution, and we choose the largest value among these two parameters.
If the chosen value is below the identification threshold, then there is neither bad data nor parameter error existing.
On the other hand, the measurement or the parameter corresponding to the chosen largest value will be identified as the error.
The part corresponding to the error will be removed, and SE and bad data detection will be carried out again.
Such procedure is performed until there is no error.

\section{Attack Model}\label{sec:attack_model}
In this section, the attacker block in Fig. \ref{fig:system_model} is illustrated.
In particular, the capabilities for the attacker and the selecting limitations of target line are first explained.
Then, the procedure of launching the proposed attack is separated into three parts for illustration which are selection of the line for attack target and decoy, determination of cyber attack region, and alteration of measurements.

\subsection{Introduction of the Attack}\label{subsec:attack_assump}
We assume that the attacker has the following capabilities:
\begin{enumerate}
\item the attacker has knowledge about the topology $\mathcal{G}$ of the entire system;
\item the attacker has the capability to observe the sub-network of $\mathcal{G}$ and perform the power flow calculation for the sub-network; and
\item the attacker has the capability to change the states of the measurements in the sub-network rather than whole network.
\end{enumerate}
To launch an attack, the attackers are limited to finite sets of target lines because of the following reasons:
\begin{enumerate}
\item the line that connects to a transformer, or in between two generators cannot be physically attacked;
\item the real and fake outage events cannot take place next to each other; otherwise, the true outage position can be easily observed if the operator goes to repair the misleading line;
\item the generator output cannot be modified;
\item the load of the buses in the attack region cannot be modified to be negative.
	  Moreover, the difference of the states and measurements before and after the attack must be controlled within a specified range; and
\item if the system is separated into two parts when a line is being attacked, then this line cannot be selected.
\end{enumerate}

\subsection{Mathematical Formulation of Selecting Attack Target Line}
To determine the lines for attack target and decoy, we employ Line Outage Distribution Factors (LODFs) matrix, denoted as $\qL \in \mathbb{R}^{{n_{br}}\times {n_{br}}}$, whose definition and calculation can be found in \cite{2009-LODF-cal}.
The $m$-th row and $n$-th column of $\qL$, $l_{m,n}$, represents the ratio of $n$-th line's power flow that will inject on $m$-th line when $n$-th line is in outage.
With LODF matrix, we can define an influence factor denoted as $\qf \in \mathbb{R}^{n_{br} \times 1}$ whose $l$-th element can be represented as
\begin{equation}\label{eq:flu_factor}
f_{l} =  \left(  \left( L_{\{ :,l\}} \right)^{T} sign(\qp) P_{l}  \right),
\end{equation}
where $L_{\{ :,l\}}$ means the $l$-th column taken from $\qL$.
The parameter $f_{l}$ shows the amount of power flow increases for the whole system, when the $l$-th line is disconnected.
Therefore, we can determine the target line to be in outage
\begin{equation}\label{eq:target_line}
l_o= \argmax_l\left\{f_l | l=1, \cdots, n_{br}\right\}.
\end{equation}
After determining the line to be disconnected, we have to choose which line is used to mislead the control center.
The idea behind misleading is to let the control center find out fake outage event in the system instead of real one so that the control center is delayed the time of detecting the real outage event and even making wrong operation or decision.
The more time the control center spends on identifying the location of real outage line, the more risk the system suffers.
Therefore, for the choice of misleading, the residual lines should reach their thermal limits as close as possible after the misleading line is disconnected.
In this context, the optimization problem of selecting the line is given as
\begin{subequations} \label{eq:mislead_opt}
\begin{align}
\mathcal{F}2 : & \max_{w_{l}, \forall l=1,\cdots, n_{br}}    \sum_{l \in \mathcal{E}  } \frac{ \overline{P}_{l} }{ P_{l}^{max} }  \label{eq:mislead_obj} \\
\mbox{s.t.}~   & w_{l} \in \{ 0,1 \},    \label{eq:select_para} \\
		       & \sum_{l }^{n_{br} } w_{l} =  1,  \label{eq:total_select} \\
		       & \overline{\qp} = \qp +  (\qw^{T} \qp) * (\qL\qw). \label{eq:power_after_select}
\end{align}
\end{subequations}
$P_{l}^{\max}$ and $\overline{P}_{l}$ denote the thermal limit and the  modified real power of $l$th line, respectively.
Equation (\ref{eq:mislead_obj}) is the objective function that sums the ratio of the flow after outage to its thermal limit for all lines.
Constraints  (\ref{eq:select_para}) and (\ref{eq:total_select}) are the equations related to the selection vector, $\qw = [w_{1} \cdots w_{n_{br}}]\in \mathbb{R}^{ n_{br} \times 1} $.
Then, the calculation of the power flow after outage based on LODF matrix is shown in Equation (\ref{eq:power_after_select}).
Therefore, the misleading line is determined as $l_m = \{~l~|~w_l \neq 0,~\forall l=1,~\cdots, n_{br} \}$.

The selected outage line, $l_{o}$, and the buses connected by $l_{o}$ are assigned to set $\mathcal{L}$. 
Meanwhile, the buses linked by the misleading line, $l_{m}$, are assigned to set $\mathcal{M}$.

\subsection{Attack Region}\label{sec:region_determine}
After selecting the target lines for physical outage and misleading, we then need to determine the attack region.
This is due to the fact that attacks should not have the ability to alter the measurements of all sensors.
Therefore, we assume that the attacks only have the limited capability that can observe and alter the sub-network of $\mathcal{G}$.
To launch the combinational attack, the attacker aims to maliciously change the measurements in a sub-network of $\mathcal{G}$ denoted as $\mathcal{\overline{G}} = \left\{ \mathcal{\overline{N}}, \mathcal{\overline{E}} \right\}$.
The buses and lines in the attack region are assigned to the set $\mathcal{\overline{N}}$ and $\mathcal{\overline{E}}$ respectively.
In the set $\mathcal{\overline{N}}$, we further separate it into two sets, $\mathcal{A}$ and $\mathcal{B}$.
The boundary buses in $\mathcal{\overline{G}}$ are assigned to the set $\mathcal{B}$ and others are placed in $\mathcal{A}$.

The key idea of finding the attack region is that we have to find a new path to re-dispatch the flow to supply the load of the buses in $\mathcal{M}$, and obtain the good estimate for power flow of $l_{o}$ and the states of the buses in $\mathcal{L}$.
The sub-network can be obtained through BFS algorithm which is detailed later.

\subsection{Measurements Modification}\label{subsec:modi_measurement}
For the measurement modification, we formulate an optimization problem taking two objectives into account.
One is to minimize the difference of measurements before and after modification due to the attacker's ability.
These measurements may contain the angles, the loads of buses, and the power flows of the lines.
However, the power flows of the lines are closely related to the angles and loads of buses, and hence the first objective can be defined as
\begin{equation}
J_{1} =  || \overline{\qp} - \qp  ||_{2},
\end{equation}
where $\overline{\qp}$ is the power flow after modification in the attack region.
Another objective is to maximize the modified measurements corresponding to the power flow at line $l_{m}$ which flows from bus $i$ to bus $j$, and can be defined as
\begin{equation}
J_{2} =   \left. \frac{ \overline{\theta}_{i} - \overline{\theta}_{j} }{x_{l}} \right|_{i, j\in \mathcal{M}, l = l_{m}} .
\end{equation}
That is, we try to prevent the amount of the flow at line $l_{m}$ from being $0$ so that it makes the attack noneffective.

We then formulate the optimization problem by considering $J_1$ and $J_2$ as follows:
\begin{subequations} \label{eq:DC_PF}
\renewcommand\minalignsep{-1em}
\begin{align}
\mathcal{F}3 :  &  \min_{ \overline{\qtheta}, \overline{\qP}^{D}, \overline{\qP} } \qquad\quad J_{1} - J_{2} &&    \label{eq:total_obj}\\
\mbox{s.t.} ~~ & \overline{\theta}_{i} = \theta_{i}, & & \forall i \in \mathcal{B}, \label{eq:theta_boundary} \\
		   & (1-\tau)\theta_{i} \leq  \overline{\theta}_{i} \leq (1+\tau)\theta_{i},   & & \forall i \in \mathcal{A}, \label{eq:theta_inside}\\
	       & (1-\tau) P_{i}^{D} \leq \overline{P}_{i}^{D} \leq (1+\tau) P_{i}^{D},  & & \forall i \in \mathcal{\overline{N}},  \label{eq:load_range} \\
		   & \overline{P}_{i}^{D}  = \sum_{l  \in \mathcal{\overline{E}} }  \overline{P}_{l} + \sum_{ l  \in \mathcal{E} \setminus 	\mathcal{\overline{E}} } P_{l},     & &  \forall i \in \mathcal{\overline{N}} , \label{eq:flow_injec}\\
	       & \overline{P}_{l} = \frac{ \overline{\theta}_{i} - \overline{\theta}_{j}  }{x_{l}}, & & \forall i, j \in \mathcal{\overline{N}}, \forall l \in \mathcal{\overline{E}},  \label{eq:flow_cal}\\
			& - P_{l}^{max} \leq \overline{P}_{l}  \leq P_{l}^{max},  & & \forall l \in \mathcal{\overline{E}}, \label{eq:line_limit}
\end{align}
\end{subequations}
where $\theta_{i}$ and $\overline{\theta}_{i}$ are the angle of bus $i$ before and after modification.
$\tau$ indicates the modification range.
Equation (\ref{eq:theta_boundary}) is that the angles of the boundary buses should remain the same, and Equation (\ref{eq:theta_inside}) shows that the changes of the buses' angle in $\mathcal{A}$ should be controlled in a range.
The load difference of bus $i$ inside the region before modification, $P_{i}^{D}$, and after modification, $\overline{P}_{i}^{D}$, should be maintained in a range shown in Equation (\ref{eq:load_range}).
Then, the power injected into the bus should meet the load as listed in Equation (\ref{eq:flow_injec}).
The midified power flow in the attack region, $\overline{P}_{l}$, is calculated by Equation (\ref{eq:flow_cal}).
In the final, Equation (\ref{eq:line_limit}) shows that the flow of the $l$-th line have to be managed under the thermal limits.

\section{Implementation Strategy}\label{sec:modi_imple}
\begin{figure}
\begin{center}
\resizebox{3.6in}{!}{%
\includegraphics*{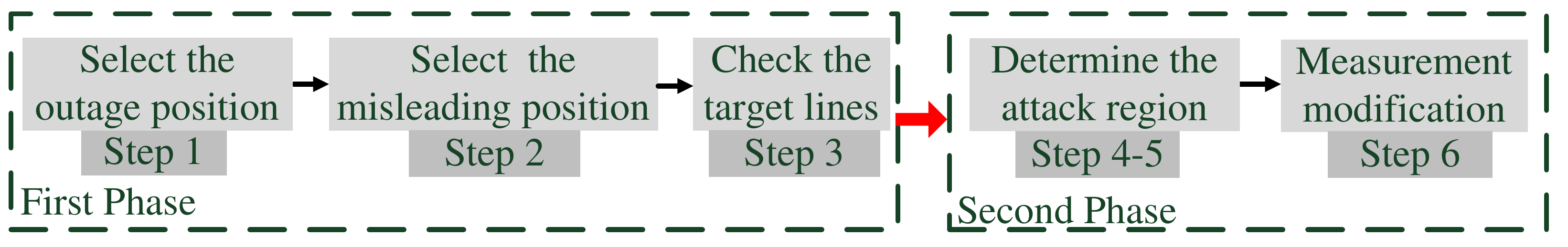} }%
\caption{The implementation strategy of the proposed attack}\label{fig:attack_procedure}
\end{center}
\end{figure}

With the description in Section \ref{sec:attack_model}, we now explain the implementation strategy of the proposed combinational attack.
The section is divided into two phases as shown in Fig. \ref{fig:attack_procedure}.
The first phase is focusing on finding the line for line-outage and misleading.
Then, with the determined lines, the attack region and the modification are illustrated in the second phase.

For the first phase, we use Equation (\ref{eq:flu_factor}) to determine the line-outage position, and then the misleading line is selected with Equation (\ref{eq:mislead_opt}).
After determining the target lines, we must check if the selection fulfills the rules described in Section \ref{subsec:attack_assump}.
The detailed steps are shown as follow:

Step 1: We wish to select the line with the greatest influence to the system for its disconnection.
Therefore, $l_{o}$ is selected as the description in (\ref{eq:target_line}), and the buses linked by $l_{o}$ are assigned to $\mathcal{L}$.
Use LODF to calculate the power flow after $l_{0}$ is disconnected.

Step 2: For the selection of the misleading, we apply misleading line select algorithm (MSLA) listed in Algorithm \ref{ago:line_sel}.
At the beginning of the algorithm, we construct a vector $\qu$.
Then, the exhaustive search is applied to calculate the objective function of (\ref{eq:mislead_opt}) which is then assigned to $\qu$.
At the same time, we have to avoid the line $l_{o}$ being selected.
In the final, the line with the largest value is selected to be $l_{m}$.

Step 3: Once we obtain the lines, we have to check if the lines are reasonable or following the rules described in Section \ref{subsec:attack_assump}.
If not, we eliminate the $l_{o}$ from $\qf$ or $l_{m}$ from $\qu$ for the unreasonable line, and then start from Step 1 again.
Otherwise, enter to the second phase.

\begin{algorithm}
\caption{Misleading Line Select Algorithm (MLSA)}
\label{ago:line_sel}
 \DontPrintSemicolon
\KwIn{Power flow $\qp$, LODF matrix $\qL$}
\KwOut{misleading line $l_{m}$}
$\qu = [u_{1} \cdots u_{n_{br}}]\in \mathbb{R}^{ n_{br} \times 1}  $. \;
\For{l  = 1 \KwTo $n_{br}$ }{
\eIf{ $l = l_{o}$ }
{
$u_{l}=0$. \;
}
{ $\overline{\qp} = \qp +  P_{l} * \qL_{ \{ 1:n_{br}, l  \} }$. \;
 $ u_{l} = \sum_{l \in \mathcal{E} \setminus  l_{o} } \frac{ \overline{P}_{l} }{ P_{l}^{max} } $.
}
$l_{m} = \argmax_{l} \left \{ u_{l} | l=1, \cdots, n_{br}  \right\}$.
   }
\end{algorithm}

In the second phase, the attack region and the modification have to be determined based on the selected lines.
The region of the sub-network $\mathcal{\overline{G}}$ is obtained by using BFS algorithm for finding the shortest path to redispatch the power flow, and the modification is based on the solution of the Problem $\mathcal{F}3$.
The detailed steps are listed as follow:

Step 4: Assuming the flow of the $l_{m}$ is from bus $i$ to bus $j$.
The trivial solution is that we just add and minus the flow amount to bus $i$ and $j$, respectively. 
However, it can be easily recognized by the control center.
To prevent from the trivial solution, we just add the flow amount to the load of bus $i$, and try to find another path to supply the load at bus $j$.

Step 5: Set the $\mathcal{\overline{N}}$ and $\mathcal{\overline{E}}$ in $\mathcal{\overline{G}}$ as empty sets first.
To find a path to supply bus $j$, we then use the BFS algorithm described in Algorithm \ref{ago:BFS_alo} to find the shortest path for redispatching the flow.
The path obtained from Algorithm \ref{ago:BFS_alo} is regarded as the sub-network $\mathcal{\overline{G}}$.
We further includes $l_{o}$ to $\mathcal{\overline{N}}$ and the buses in $\mathcal{L}$ to $\mathcal{\overline{E}}$ as the attack region.

Step 6: With the attack region, we now solve the optimization Problem $\mathcal{F}3$.
The formulation in (\ref{eq:DC_PF}) is a convex optimization problem with linear constraints.
There are many existing algorithms and toolboxes dealing with  convex optimization problem; therefore, one of them is applied.
If the Problem $\mathcal{F}3$ has no solution, which means the current attack region cannot satisfy the constraints.
Algorithm \ref{ago:BFS_alo} is thus applied again, and go back to solve Problem $\mathcal{F}3$ again.
With the solutions, set $\overline{\qz} = \qz$ and replace the measurements of $\overline{\qz}$ in $\overline{\mathcal{G}}$ with the solution of Problem $\mathcal{F}3$.

\begin{algorithm}
\caption{BFS algorithm for finding misleading line}
\label{ago:BFS_alo}
 \DontPrintSemicolon
\KwIn{System topology $\mathcal{G}$, bus $j$, sub-network $\mathcal{\overline{G}}$, line $l_{o}$}
\KwOut{Sub-network $\mathcal{\overline{G}}$}
Find a bus $g$ which has a generator and is the nearest to bus $j$. \;
Current system configuration is $\mathcal{W} = \left\{ \mathcal{N}, \mathcal{E}\setminus \left\{ \mathcal{\overline{E}}, l_{o} \right\} \right\}$. \;
$g$ : starting bus.   $j$ : destination bus.  \;
let the bus $g$ be the {\it progress bus} and the level $k=0$.
Rest buses are set as {\it unvisited buses}.\;
Search all of the {\it unvisited buses} connected to the buses in {\it progress buses}.
Put such {\it unvisited buses} to {\it progress buses} and previous {\it progress buses} are assigned as {\it visited buses}.
\;
\eIf{ $j \in$ {\it progress bus} }
{
go to step 11 of Algorithm \ref{ago:BFS_alo}. \;
}
{repeat step $5$ of Algorithm \ref{ago:BFS_alo} again.\;
$k=k+1$. \;
}
Backtrack from the destination bus to the starting bus level-by-Ievel, and identify the shortest path.
The buses and lines in the path are given to $\mathcal{\overline{N}}$ and $\mathcal{\overline{E}}$ respectively.\;
\end{algorithm}

\section{Case Study}\label{sec:simulation}
\begin{figure}
\begin{center}
\resizebox{2in}{!}{%
\includegraphics*{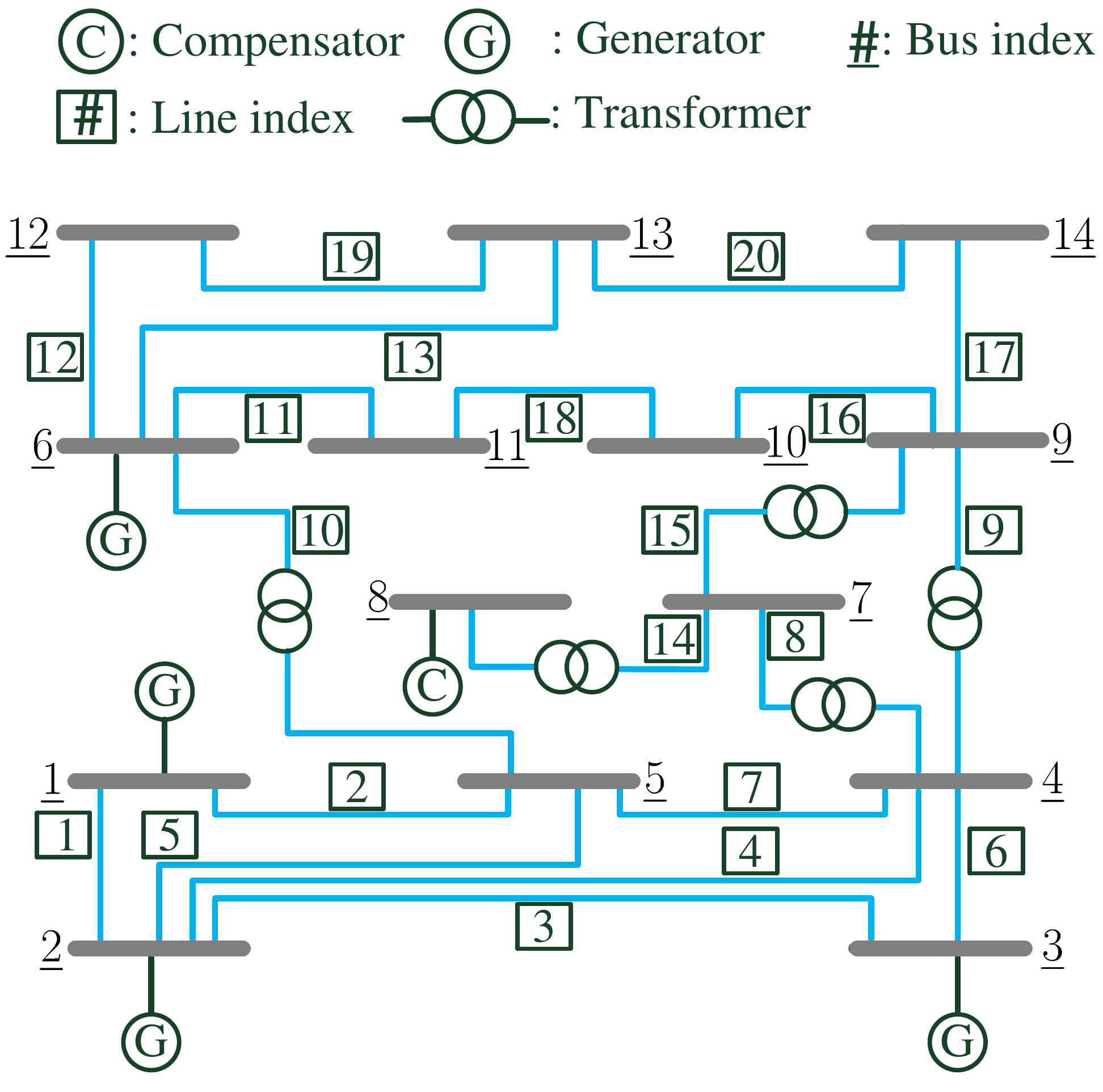} }%
\caption{IEEE 14-bus test system \cite{IEEE_14bus}} \label{fig:14bus_system}
\end{center}
\end{figure}

In this section, we adopt the IEEE 14-bus system \cite{IEEE_14bus} to illustrate the proposed attacking mechanism in detail.
The system topology is shown in Fig. \ref{fig:14bus_system}, and the thermal limit of each line is listed in Table \ref{tb:flow_limit}.
Without any specification, the modification rage, $\tau$, for all measurements is set to $50 \%$.
The errors for all measurements are assumed to be $n_{i} \sim N(0, 0.001)$.
The identification threshold of bad data and parameter error detection is set to $2$ which is outside of $95 \%$ confidence interval.
The software toolbox, MATPOWER \cite{MATPOWER}, is utilized to run the power flow to provide the initial information of the system.
To solve the Problem $\mathcal{F}3$, we use CVX \cite{CVX}, a package intended to solve convex programs.

\begin{table} \scriptsize
\begin{center}
\caption{The thermal flow limit of IEEE 14-bus system}\label{tb:flow_limit}
\begin{tabular}{|c|c|c|c|c|c|c|c|c|c|}
\hline
Line    & limit     & Line    	& limit		& Line    	& limit  	& Line    	& limit		  \\
number  & (MW)      & number    & (MW)      & number 	& (MW) 		& number 	& (MW)\\
\hline
$1$    & $200$      & $6$    	& $50$      & $11$    	& $50$  	& $16$    	& $20$  \\
\hline
$2$    & $100$      & $7$    	& $100$  	& $12$    	& $20$  	& $17$    	& $20$ \\
\hline
$3$    & $100$      & $8$    	& $50$   	& $13$    	& $50$  	& $18$    	& $20$  \\
\hline
$4$    & $100$      & $9$    	& $100$   	& $14$    	& $50$  	& $19$    	& $20$  \\
\hline
$5$    & $100$      & $10$    	& $100$  	& $15$   	& $50$  	& $20$    	& $20$  \\
\hline
\end{tabular}
\end{center}
\end{table}

In the beginning, we select the target line based on the Step 1 to Step 3 in Section \ref{sec:modi_imple}.
Line $3$ is first selected for line outage and line $10$ is the line for misleading.
However, the line $3$ connects two generators, and there is a transformer on line $10$; hence, we have to choose the target lines again.
Following the proposed recursive way, the line $13$ and $17$ are finally selected as the line for outage and misleading respectively.

The direction of the misleading line is from bus $9$ to bus $14$ so that we have to find the path to supply the load of bus $14$.
Moreover, the nearest generator is at bus $6$.
Therefore, we now use the Algorithm \ref{ago:BFS_alo} to find the shortest path from the starting bus, bus $6$, to the destination bus, bus $14$.
Table \ref{tb:set_detail} summarizes the attack region based on the results of Algorithm \ref{ago:BFS_alo}.
Then, the measurements before and after modification based on the results of Problem $\mathcal{F}3$ are listed in Table \ref{tb:modi_load_phase} and \ref{tb:modi_flow}.

\begin{table} \footnotesize 
\begin{center}
\caption{The description of the sets used in the modification}\label{tb:set_detail}
\begin{tabular}{|c|c|c|c|c|c|c|c|c|}
\hline
Set   	&  Bus number		& Description  \\
\hline
$\mathcal{A}$    & $12, 13, 14$  & The buses in the attack region          \\
\hline
$\mathcal{B}$    & $6$           & The boundary bus of the attack region \\
\hline
$\mathcal{L}$     & $6, 13$      & The buses connecting the line-outage line \\
\hline
$\mathcal{M}$    & $9, 14$       & The buses connecting the misleading line       \\
\hline
\hline
Set   	&  Line number		&　Description  \\
\hline
$\mathcal{\overline{E}}$    & $12, 13, 19, 20$       & The lines in the attack region      \\
\hline
\end{tabular}
\end{center}
\end{table}

\begin{table} \footnotesize 
\begin{center}
\caption{The phase and load before and after modification}\label{tb:modi_load_phase}
\begin{tabular}{|c|c|c|c|c|c|c|c|c|}
\hline
Bus number   	& \multicolumn{2}{c|}{Phase (angle)} 		  &  \multicolumn{2}{c|}{Load (MW)} \\
\hline
	  & Before & After & Before & After\\
\hline
$6$     & $-0.1378$       & $-0.1378$        & $11.20$ 		& $ 9.21 $    \\
\hline
$9$     & $-0.1615$       & $-0.1615$        & $29.50$      & $ 37.89 $   \\
\hline
$12$    & $-0.1582$       & $-0.1592$        & $6.10$ 		& $ 6.20 $    \\
\hline
$13$    & $-0.1616$       & $-0.1635$        & $13.50$      & $ 12.13 $    \\
\hline
$14$    & $-0.1842$       & $-0.1975$        & $14.90$      & $ 9.76 $   \\
\hline

\end{tabular}
\end{center}
\end{table}

With the modified measurements, we perform SE and then bad data and parameter error detection.
The equation of the power flow is linear, the solution of SE can be easily obtained as
\begin{equation}
\widehat{\qtheta} = \left( \qB_{f}^{T} \qR^{-1} \qB_{f} \right)^{-1} \qB_{f}^{T} \qR^{-1} \overline{\qz}.
\end{equation}
Hence, we apply the detection by calculating the normalized residual and parameter errors, and sort the results shown in Table \ref{tb:bad_data_detec}(a) in a descending order.
From the table, there are two largest parameter errors related to $x_{17}$ and $x_{20}$ and they are also larger than the identification threshold.
We then eliminate the measurements having relation with $x_{17}$ and $x_{20}$, and apply the bad data detection again.
Table \ref{tb:bad_data_detec}(b) shows the results of the second-round detection.
The largest value in Table \ref{tb:bad_data_detec}(b) is much lower than the threshold.
Therefore, according to the results, we successfully let the control center find out there is an error happening in the misleading line.

The bad data and parameter error detection can be influenced by the noise, we further collect the results with $1,000$ Monte Carlo simulations.
If the parameter of misleading line is recognized as the error, and the corresponding parameter error is larger than the threshold, the attack is regarded as a successful attack.
Furthermore, the false alarm is defined as other parameter or measurement are regarded as the error.
According to the results, the successful rate calculated by the ratio of the number of the successful attacks to $1,000$ is $79.90\%$, and the false alarm rate is $0\%$.
That is, the error of the parameter at misleading line shows up at every simulation. 
However, the normalized parameter errors are sometimes not larger than the threshold with noise's influence.
Therefore, we can ensure the efficiency of the proposed attack strategy.

\begin{table} \footnotesize 
\begin{center}
\caption{The power flow before and after modification}\label{tb:modi_flow}
\begin{tabular}{|c|c|c|c|c|c|c|c|c|}
\hline
Line number   	& \multicolumn{2}{c|}{Power flow (MW)} 		  \\
\hline
	  & Before & After\\
\hline
$12$    & $7.88$       & $8.36$      \\
\hline
$13$    & $18.22$      & $19.36$      \\
\hline
$17$    & $8.39$       & $0$         \\
\hline
$19$    & $1.78$       & $2.16$         \\
\hline
$20$    & $6.51$       & $9.76$         \\
\hline

\end{tabular}
\end{center}
\end{table}

\begin{table} \footnotesize 
\begin{center}
\caption{The bad data and parameter error detection results}\label{tb:bad_data_detec}
\begin{tabular}{|c|c|c|c|c|c|c|}
   \multicolumn{2}{c}{(a) First Round}                                                          &  \multicolumn{3}{c}{ ~~~~(b) Second Round}     \\
   \cline{1-2} \cline{4-5}
   \multirow{1}{*}{Parameter}   & \multirow{2}{*}{$\lambda_{i}^{N}$, $r_{i}^{N}$ }  & & \multirow{1}{*}{Parameter}   & \multirow{2}{*}{$\lambda_{i}^{N}$, $r_{i}^{N}$ } \\
   \multirow{1}{*}{Measurement} &                                                   & & \multirow{1}{*}{Measurement} &　\\ \cline{1-2} \cline{4-5}   \cline{1-2} \cline{4-5}
   $x_{17}, x_{20}$ 			& $3.0126$                                          & & $x_{3}, x_{6}$    & $0.0531$ \\    \cline{1-2} \cline{4-5}
   $x_{13}$						& $1.4691$                                          & & $x_{5}$           & $0.0339$ \\    \cline{1-2} \cline{4-5}
   $x_{11}, x_{16}, x_{18}$		& $1.3031$                                          & & $x_{7}$           & $0.0259$ \\    \cline{1-2} \cline{4-5}
   $x_{10}$						& $1.0580$                                          & & $x_{9}$           & $0.0182$ \\    \cline{1-2} \cline{4-5}
 \cline{1-2} \cline{4-5}
\end{tabular}
\end{center}
\end{table}

\section{Conclusions}\label{sec:conclusion}
In this paper, we present the combinational attack which maliciously injects the false data in the cyber layer to cover the physical event in the power system.
While launching the attack, the method of finding the target lines is introduced based on the LODF matrix.
Moreover, an algorithm followed by BFS algorithm was proposed to find the attack region, and the modification results are from the power flow equations.
The simulation results also reveal that the proposed scheme can successfully achieve the goal of misleading the control center and mask the line-outage event.
As the future work, we will extend this study in two directions by proposing the attack based on AC power flow, and investigating a protection strategy for the cyber-physical system.
Moreover, the assessment of the power system with the proposed attack method should also be discussed.

\section{Acknowledgements}
This work was supported by Ministry of Science and Technology under grant numbers MOST 105-2221-E-001-009-MY3 and 104-2221-E-001-008-MY3, and Academia Sinica Thematic Project AS-104-TP-A05.

\bibliography{IEEEabrv,references}

\end{document}